\documentclass[prb,aps,amssymb,amsmath,twocolumn]{revtex4}
\usepackage{epsfig}
\begin{document}
\renewcommand{\thefootnote}{\fnsymbol{footnote}} 

\title{Prospects of Transition Interface Sampling simulations for the theoretical study of zeolite synthesis}

\author{Titus S. van Erp}
\email[]{E-mail: Titus.VanErp@biw.kuleuven.be}
\author{Tom P. Caremans}
\author{Christine E. A. Kirschhock} 
\author{Johan A. Martens}

\affiliation{Centre for Surface Chemistry and Catalysis, 
K.U. Leuven, Kasteelpark Arenberg 23, 3001 Leuven, Belgium} 
\date{\today}

\begin{abstract}
\noindent
The transition interface sampling (TIS) technique
allows to overcome large free energy barriers within reasonable
simulation time,
which is impossible for straightforward molecular dynamics. Still,
the method does not impose an artificial driving force, but it
surmounts the timescale problem by an importance sampling of true dynamical 
pathways. Recently, it was shown that the efficiency of TIS 
to calculate reaction rates
is less sensitive to the choice of reaction coordinate than those of 
the standard free 
energy based techniques. This could be an important 
advantage in complex systems for which a good reaction coordinate is     
usually very difficult to find.
We explain the principles of this method and discuss
some of the promising applications related to zeolite formation.
 
\end{abstract}

\maketitle
\section{Introduction}
Gaining insight in the zeolite formation has not only fundamental scientific 
importance, but could also accelerate momentous technological developments.
The applications of zeolites are uncountable ranging from cracking catalysis
of crude oil, gas separation, detergent builders, and sensors
for pharmaceutical formulations.
The specific catalytic properties of zeolites lie in their unique open 
crystalline structure that incorporates cages or channels with typically
nanoscale diameters. The growth of the open structure silicon dioxide
polymorphs 
is mediated by 
so-called template molecules that can be removed out of the zeolite pores
after the crystallization process.
Besides template molecules, solvent, Si/Al ratio, temperature, pH, 
and many other
factors play a role in determining which zeolite topology is formed. 
As each structure and composition
has its unique catalytic properties,
the synthesis of new zeolite materials
has been an important branch of chemical research.
This development has progressed mainly on the basis of 
trial-and-error and 'chemical intuition' as a fundamental understanding of
zeolite formation is lacking. 
The clear solution synthesis studies of silicalite-1 zeolites initiated by
Schoeman \emph{et al.}\cite{schoeman94} were an important step forward
for the experimental analysis. 
The use of clear solutions instead of gels made the analysis 
of zeolite synthesis
much more accessible by experimental techniques. Since then, this model system 
has been subject of many studies including
x-ray and neutron scattering, infrared (IR) spectroscopy, 
Nuclear magnetic resonance (NMR), 
and dynamic light scattering (DLS).
These studies revealed that upon mixing tetraethylorthosilicate (TEOS),
tetrapropylammonium hydroxide (TPAOH) and water at a certain ratio 
at room temperature 
sub-colloidal particles are formed of several nanometers.
Using the freeze drying technique~\cite{schoeman96,freeze2},  
these particles have been extracted from 
the solution and examined by various techniques such as
solid state NMR, Fourier Transform IR (FTIR), 
transmission electron microscopy (TEM), 
and atomic force microscopy (AFM). Various models for the structure of these
particles have been proposed ranging from amorphous 
bodies~\cite{kragten03,Rimer06} to precise framework 
fragments~\cite{nanoslab1}.

The formation of crystalline zeolite particles is initiated when this 
suspension
is heated upto temperatures of 350 K. 
Light scattering experiments show that the intensity scattered by the suspension increases only slowly in time during the first period of the synthesis. This is then followed by a sharp increase, indicating the starting point of growth of what will become the final crystals.
The first period
can be associated to a nucleation process, in which a particle has to be
formed with a size beyond its critical nucleation radius. 
The formation of zeolites consist hence of several stages. First a 
polymerization process which eventually leads to the formation of 
sub-colloidal particles, second the nucleation process, and finally the
crystal growth.

One of the difficulties in the 
investigation 
of the zeolite formation
process is that the relevant lengthscales of the
zeolite formation lie just in between the accessible lengthscale of NMR and
diffraction techniques~\cite{auerbach05}. Moreover, 
it is unclear if freezed-dried extractions are identical 
to the silicate particles existing in solution.
Since many experiments do not  
allow unequivocal interpretation, it is not a surprise that several
crystallization mechanisms 
have been proposed. These theories concentrate on the structure and shape 
of the
colloidal particles, how these particles are formed and how these particles
finally contribute in the formation of the zeolite crystal.

An example is the nanoslab hypothesis that was postulated by some of 
us. 
It was inspired by several experimental 
observations~\cite{nanoslab1,nanoslab2,nanoslab3}. This 
theory assumes that at an initial stage precursor particles are formed that
consist of 30 to 33 Si atoms enclosing a single template molecule. These 
precursors stick together in a block shaped particle, the nanoslab, 
that has already the correct crystalline structure.  These particles
finally form the 
zeolite by a 'clicking-mechanism' 
when the solution is heated up.
Others have claimed that the apparent evidence of the Si-30/33 precursor particle
should be attributed to other Si containing species~\cite{knight06} or that
the nanoshaped particle is actually an amorphous identity with a layer 
of template molecules around it~\cite{kragten03,Rimer06}. Also
the role of the nanosized particles for the nucleation and crystal growth 
has been subject of 
debate. 
According to some groups, the nanoparticles add one by one to the growing 
crystals~\cite{niko2000,dokter}. Others regard the particles as monomer 
reservoirs: monomer dissolves into the solution and attaches to the growing 
nucleus~\cite{Schoeman98,cundy}. Recent publications~\cite{davis,cundy}  
state that an aggregative growth mechanism of discrete nanoparticles 
may dominate the early stage of the growth process. However, after 
a certain size is attained, the growth mechanism seems to switch 
to addition of low molecular species, probably monomers.

In conclusion, despite many years of abundant experimental research, 
zeolite synthesis still contains many mysteries. Therefore, this 
field of research is a prototype example where computer simulations
could give invaluable information. However, before truly realistic simulations
of all stages in  the zeolite synthesis can be performed, a long way has
to be gone. Reason for the difficulty is that the typical system sizes and 
timescales
at which the zeolite formation takes place are generally beyond the 
capabilities
of present computer resources. For a correct modeling of the nucleation 
process,
the simulation box should at least be larger than the critical nucleus. 
A requirement 
that is out of limits for quantum mechanical calculations and demands the 
development
of accurate reactive forcefields.  

So far fully quantum mechanical 
calculations using Density
Functional Theory (DFT) have been applied to silica polymerization 
clusters~\cite{Pereira1,Pereira2}.
These studies showed a stronger stability of silicate 6 rings 
and linear polymers compared to smaller rings and branched polymers.
Ph effects were considered in [\onlinecite{Lewis2,Thuat}] by
analyzing negatively charged silica clusters that are 
favorable to neutral ones in an alkaline environment. This study revealed
that internal cyclization if preferred over further linear growth~\cite{Lewis2}.
Barriers for oligomerization were significantly reduced for single charged 
cluster compared to neutral ones~\cite{Thuat}. 

Based on ab initio calculations or experimental data, several classical 
forcefields have been 
developed~\cite{catlow2, aoki,vashista,beest, feuston}. These potentials 
allow the
study of larger systems
including solvent and template molecules. However, the existing potentials 
are not yet very accurately
describing the breaking and making of chemical bonds, which 
presumably requires complex many-body terms and polarizable
forcefields. Still, studies using these approximate potentials can give 
valuable insights.
For instance, classical molecular dynamics (MD) simulations have shed 
some light to the role of
solvent and template molecules~\cite{Catlow1,Lewis}. These simulations
showed that, contrary to  fully formed cages and rings, open structures 
collapse in the presence of solvent, unless it contained
strongly bonded template molecules. 
The early stages of silica polymerization
dynamics were studied by Rao and Gelb~\cite{Rao} at high temperatures 
$\gtrsim 1500$ K.
These alleviated temperatures were required as upto $600$ K, 
no polymerization reaction could be observed within
the nanoseconds simulation periods. 
They found that both the monomer incorporation and the 
cluster-cluster aggregation were important mechanisms for diluted solutions,
while the first mechanism was dominant in the concentrated systems.
Using a implicit solvent model not including template molecules, 
Wu and Deem analyzed the free energy barriers and 
critical cluster sizes as function of pH and Si-monomer concentration 
at ambient conditions using a series of advanced Monte Carlo (MC) 
techniques~\cite{Deem1}. They found that the critical clusters 
for the polymerization contained relatively few ($\approx 30-40$) Si atoms.
No attempt was made to derive 
reaction rates by calculating transmission coefficients. 
Even larger systems and timescales have been simulated using lattice 
models~\cite{auerbach05jacs,Gubbins}
and kinetic MC (KMC)~\cite{G78}. 
Relative rates for different crystal growth mechanisms
via kink and edge sites can 
be derived by mimicking atomic force micrographs via atomistic 
simulations~\cite{agger1,agger2,agger3}. 
Still, even KMC 
simulations are usually restricted to 
growth~\cite{gale1,gale2}. The time, before the critical nucleus 
of a zeolite
is formed, is still too long 
even for this ultrafast type of dynamical simulations. 

It is clear that the simulation methods have made significant progress 
in recent years. 
At the early stages
of Si polymerization,
fully quantum mechanical  MD studies are in our reach using 
Born-Oppenheimer~\cite{marx} or
Car-Parrinello~\cite{marx,cp} simulations.
Thanks to newly developed potentials and coarse grained systems,
simulations approach the system sizes that are needed
to describe the template 
directed zeolite synthesis in solution. 
Nonetheless, each stage in the zeolite synthesis involves significant 
reaction barriers. 
This makes the chance to observe important reactive events 
at experimental conditions
within the duration of the simulation period highly unlikely.
The reaction itself is usually very fast and could fit perfectly
within the window of timescales that are attainable 
by the simulation method. However, the system will likely spend extensively 
long periods within the well of the reactant state without 
any reactive event taking place. 
It is, therefore, important to have a method that focuses the costly 
simulation time on the important but rare reactive events,
while limiting the superfluent exploration of the reactant well. 
In this article,
we review such a method, the transition interface sampling 
(TIS)~\cite{ErpMoBol2003,titusthesis,ErpBol2004} method, that allows 
to concentrate only on those trajectories that are important for the 
chemical process. Moreover, the TIS technique can calculate
the  frequency for occurrence of these successful trajectories 
within an infinitely 
long straightforward simulation. 
Hence, TIS allows the determination of the rate of the rare event.  

The aim of this article is not to give a fully
detailed theoretical derivation of the method. 
This has already been published 
elsewhere~\cite{ErpMoBol2003,titusthesis,ErpBol2004}. The goal of this article
is to give an educative overview of the practical
algorithms and their possible applications
related to zeolite studies
 rather than on 
mathematical aspects.
  
\section{Transition Interface Sampling}
\subsection{historic perspectives of rare event simulations}
The first theories for treating rare events from a microscopic 
perspective where pioneered
by Eyring~\cite{E35}, Wigner~\cite{W38}, and Horiutu~\cite{H38} 
about 20 years before the first MD simulation
was performed~\cite{Alder}. They introduced the concept of 
Transition State (TS) and the so-called 
TS Theory (TST) approximation. Later Keck demonstrated how the 
TST approximation can be made exact 
by a dynamical correction, the transmission coefficient~\cite{Keck67}.
The actual application of these  theories for molecular simulation 
was directed by the works of
Bennett~\cite{Bennet77} and Chandler~\cite{DC78}, which have made this a 
standard approach in molecular simulation.
A crucial point in this reactive flux (RF) method is the definition of a 
suitable reaction coordinate (RC).
As a first step, the free energy needs to be determined along this RC using 
importance sampling techniques such
as Umbrella Sampling (US)~\cite{TV74} or 
Thermodynamic Integration (TI)~\cite{CCH89}. This result alone 
is sufficient to obtain the TST approximation of the rate, which is an upper 
limit for the actual rate. 
In the second step, the correction to 
this approximation can be calculated by releasing dynamical trajectories from 
the top of the free energy barrier.
Only when both steps are completed, the exact reaction rate can be calculated. 
Both the free energy barrier
and the transmission coefficient depend on which RC  is taken, but the 
final result that 
combines the two outcomes is independent of this choice. 

The RF method as proved its value for many systems, but also has its drawbacks. 
Although its result is independent
of the chosen RC, its efficiency does and sensitively determines its success 
or failure.
A non-suitable choice of RC can result in hysteresis effects in the free energy 
calculation, which
frustrates an accurate estimation of the barrier. Besides, even if an accurate 
value for the free energy barrier can be obtained,  the 
corresponding transmission coefficient will be 
very small and its evaluation will 
require an extremely 
large number of pathways. In practice, it has been experienced that finding 
a good RC can be extremely difficult in high dimensional complex systems.
Notable examples are chemical reactions in solution, where the 
reaction mechanism often depends on  highly non-trivial solvent
rearrangements.
Also, computer simulations of nucleation processes use very complicated  
order parameters to 
distinguish between particles belonging to the liquid and solid phase. 
This makes it unfeasible to construct a single RC that accurately describes 
the exact place of cross-over transitions.  
As result, hysteresis effects and low transmission coefficients are almost 
unavoidable.

The problem of finding suitable RCs, has urged the development of 
alternative methods. In 1998, 
Dellago \emph{et al.} came up with such an alternative method that they called 
transition path sampling (TPS)~\cite{TPS98,TPS98_2,TPS98_3,TPS99}. 
This approach can 
be described as a MC sampling of MD pathways.
Using a detailed balance technique,  a set of trajectories can be collected 
that satisfy some predetermined
criteria. For instance, one can constrain the start- and end-point 
of the path in such a way that each trajectory
connects the reactant and product state.  An important point is that this 
sampling of successful reactive events
does not require a RC that captures the reactive mechanism, but only needs 
an order parameter that can distinguish 
between reactant and product state. In addition to this, the first series of 
TPS papers~\cite{TPS98,TPS98_2,TPS98_3,TPS99} also provided 
a route to 
calculate reaction rates.
However, this approach has seldom been used due to its 
high computational cost.  
Moreover, within the context of the reaction rate calculation, it is not so
obvious to state that
the TPS order parameter is actually very different from a RC.
In this approach, the end-point of the path is forced to progress
in successive steps from reactant to product state. Hence, the TPS order
parameter needs to describe the intermediate states as well just as
a RC in the standard methods. 

Luckily, the algorithmic procedure  to calculate reaction 
rates using the  same path sampling 
framework
improved considerably when the 
transition interface sampling (TIS) technique was devised~\cite{ErpMoBol2003}. 
TIS uses a flexible pathlength which reduces the number of required MD steps
significantly. Moreover, 
the TIS method also eliminates the need of so-called MC \emph{shifting} moves
that required a considerable percentage of the simulation time in the 
TPS scheme. 
In addition, one can show that the new mathematical formulation 
of the reaction
rate is less sensitive
to recrossing events which guarantees a faster convergence. 

While TPS imposes conditions to the start- and
end-point of the path to be within certain intervals of the RC, TIS 
imposes an interface crossing condition. 
Except for the technique proposed in Ref.~[\onlinecite{ErpBol2004}], TIS needs 
a RC just like the original TPS scheme. The RC is required to define a set of 
interfaces between the stable reactant and product states.
However, unlike the standard RF methods, the TIS efficiency is relatively
insensitive to the choice of RC as was first proven 
in~[\onlinecite{TISeff}]. This point is a strong advantage 
in complex systems
where a 'good RC' can be extremely difficult to find.
\begin{figure*}[ht!]
  \begin{center}
  \includegraphics[width=15.0cm]{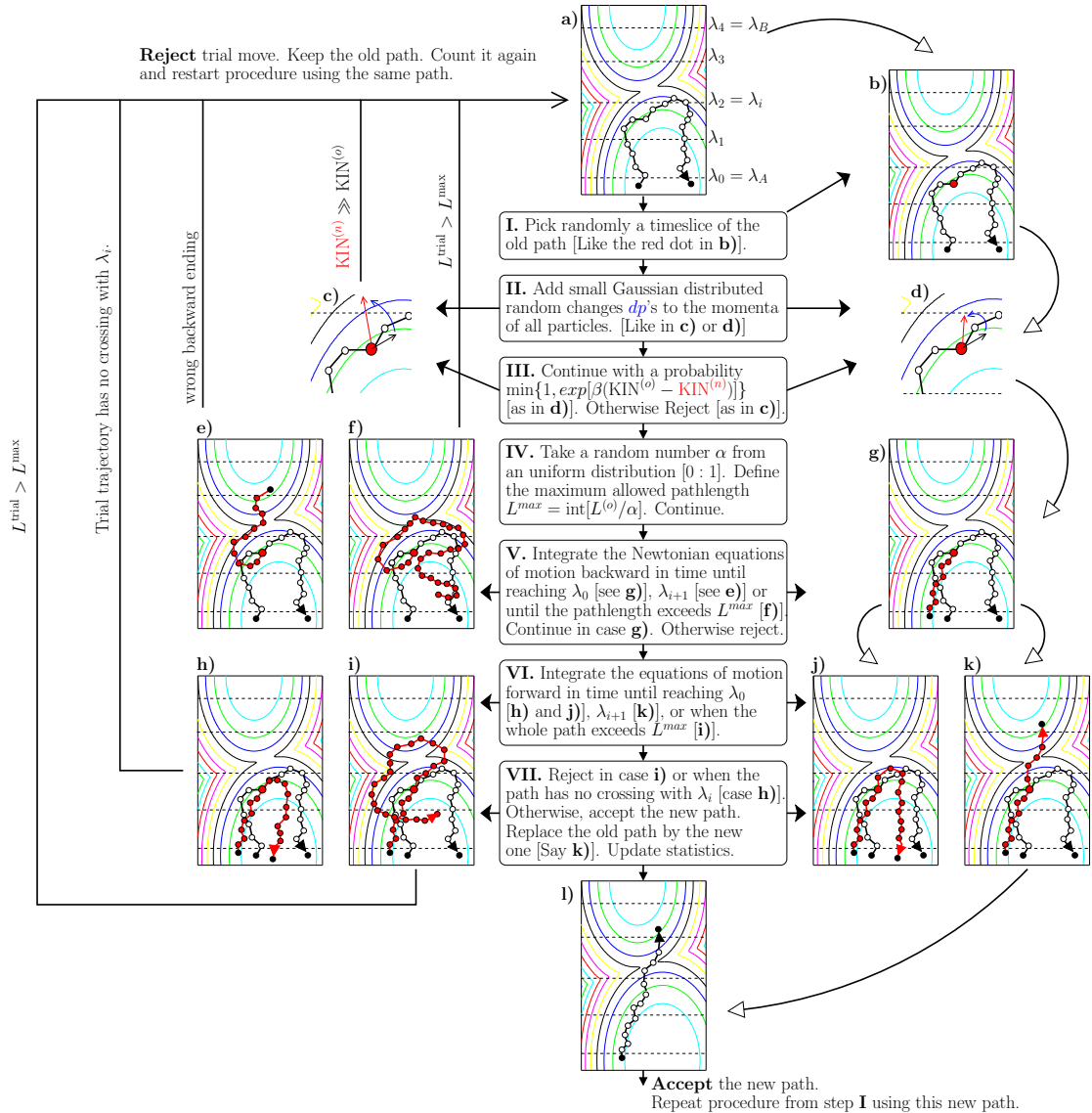}
   \caption{Illustration of the shooting algorithm in TIS.
The panels {\bf a)}-{\bf l)} depict
trajectories/trajectory-segments on a free energy surface.
The dashed horizontal lines
are the TIS interfaces ($n=4$ and $i=2$ in this case).
The algorithm requires an initial
path {\bf a)} to start the loop. The length of this particular
path $L^{(o)}$ is eighteen timeslices (endpoints are
not included).  At step {\bf I}, a random point is picked from this old path
and some small randomized
changes are applied to the velocities of all the particles
({\bf II}), followed by a
  Metropolis acceptance/rejection step ({\bf III}).
In {\bf c)} the new velocities have resulted in a
much larger
kinetic energy ${\rm KIN}^{(n)}$
  and therefore this trial move is most likely rejected. Step {\bf IV} is requir
ed
to maintain detailed balance
  between pathways of different lengths. For example, if the random number
generator
  assigns $\alpha=.59$ then $L^{\rm max}=30$ and we can reject when the path
is unfinished, but already contains 31 timeslices as in panel {\bf f)}
and {\bf i)}. At {\bf V}, the equations of motion
  are integrated
  backwards in time by a MD algorithm using the shooting point with
reverse velocities as starting point.
At {\bf VI)} the equations of motion are integrated forward in time
starting from the same shooting
point (without reversed velocities).  After a rejection the old path is
kept and counted again.
If accepted, the new path will automatically start at $\lambda_A$ and
cross $\lambda_i$. The path
can end at either $\lambda_{A}$ as in {\bf j)} or at $\lambda_{i+1}$ as
in {\bf k)}.
The fraction of sampled pathways that end at $\lambda_{i+1}$
determines ${\mathcal P}_A(\lambda_{i+1}|
 \lambda_i)$
\label{shoot}}
  \end{center}
\end{figure*}

\subsection{the TIS algorithm}
The TIS algorithm works as follows. First step is to define a RC 
and a set of related values $\lambda_0, \lambda_1, \ldots, \lambda_n$
with $\lambda_i < \lambda_{i+1}$. The subsets of phase- or configuration
points for which the RC is exactly equal to $\lambda_i$ basically define
multidimensional surfaces or interfaces which give the name 
to this method. These values/interfaces should obey the following requirements:
if the RC is lower than $\lambda_0=\lambda_A$, the system should 
be in the reactant 
state $A$; if the RC is higher than $\lambda_n=\lambda_B$ the system should be
in the product state $B$; $n$ and the positions for the interfaces in between, 
$\lambda_i$ with $1=1,2,\ldots,n-1$, should be set to optimize the efficiency. 
Further, the surface $\lambda_A$ should be set in such a way that whenever a MD 
simulation is released from within the reactant well, this 
surface should be frequently crossed. The TIS rate expression can then be
formulated as
\begin{align}
k_{AB}=f_A \, {\mathcal P}_A(\lambda_B|\lambda_A)
\label{kTIS}
\end{align}
Here, $f_A$ is the escape flux through the first interface. 
In a long MD simulation,
this simply corresponds to the number of detected crossings  
through the surface $\lambda_A$
divided by the total simulation time 
(here, we assume that we will not observe a spontaneous transition to state $B$
during the simulation.
For a more formal definition see [\onlinecite{ErpMoBol2003}]).
The other term ${\mathcal P}_A(\lambda_B|\lambda_A)$ is 
the overall crossing probability.  
This is the probability
that whenever the system crosses $\lambda_A$, it will cross $\lambda_B$
\emph{before} it crosses $\lambda_A$ again. As $\lambda_B$ is a surface at the 
other side of the barrier, this probability will be very small and
can not be calculated directly.
This probability can, however, be determined by a series of path sampling
simulations using the following factorization:
\begin{align}
{\mathcal P}_A(\lambda_B|\lambda_A)={\mathcal P}_A(\lambda_n|\lambda_0)=
\prod_{i=0}^{n-1} {\mathcal P}_A(\lambda_{i+1}|\lambda_i).
\label{kTIS2}
\end{align}
The terms ${\mathcal P}_A(\lambda_{i+1}|\lambda_i)$ 
are history  dependent
conditional crossing probabilities which are much higher and can be computed.
In words, ${\mathcal P}_A(\lambda_{i+1}|\lambda_i)$ is 
the probability that $\lambda_{i+1}$ will be crossed before $\lambda_A$
under the twofold condition that
the system is at the point to cross the interface
$\lambda_i$ in one timestep while  $\lambda_A$ was more recently crossed
than $\lambda_i$ in the past. It is due to this history dependence
that Eq.~(\ref{kTIS2}) is exact and should not be misinterpreted as
a Markovian approximation.
${\mathcal P}_A(\lambda_{i+1}|\lambda_i)$
is also equal to the number of all possible paths that start at $\lambda_A$
and end at $\lambda_{i+1}$ divided by the number of all possible paths
that start at $\lambda_A$, end at either $\lambda_{i+1}$ or $\lambda_{A}$,
and have at least one crossing with $\lambda_i$. 
Hence, this term can be calculated if we can generate the appropriate 
trajectories with their correct statistical weight. 
It is however not so
obvious to generate these pathways especially when $\lambda_i$ 
is in the reaction barrier region. This difficulty can be overcome
by a MC algorithm that employs a variation of the TPS \emph{shooting} move.
The algorithm is explained in Fig.~\ref{shoot}.

The  algorithm requires to have one path
fulfilling the correct condition. 
That is starting
 at $\lambda_A$ and crossing $\lambda_i$ at least once before ending 
at either $\lambda_{i+1}$ or $\lambda_A$.
A crucial point is step II. After picking a random 
timeslice (a point that constitutes
all the particle positions and momenta at a certain timestep along the path), 
one adds random values 
to all the momenta. In practice, these random values are taken from a 
Gaussian distribution
with a certain width $\sigma$, that should be adapted to obtain the 
optimum efficiency.
If $\sigma$ is small, the random momentum changes will be small as well 
and the new path will lie 
closely to the old one (if we assume deterministic dynamics). 
This small deviation results in a significant 
chance that the trial path  will satisfy the required conditions as well 
which yields a good 
acceptance rate. However, a too small value of $\sigma$ will result in 
too strong correlations
between the accepted moves (In the extreme case when $\sigma=0$, 
one regenerates exclusively the same path). Usually, one tries several values for $\sigma$ in a series 
of short test simulations.
It is generally assumed that the value that yields an acceptance rate 
of 50 \% is close to an optimum
value for $\sigma$. 
If one wants to simulate at constant energy instead of constant temperature,
step III can be replaced by a proper velocity rescaling procedure that,
if needed,
can also preserve linear and angular momentum~\cite{GDC99_2}.

\begin{figure}[ht!]
  \begin{center}
  \includegraphics[width=7.1cm]{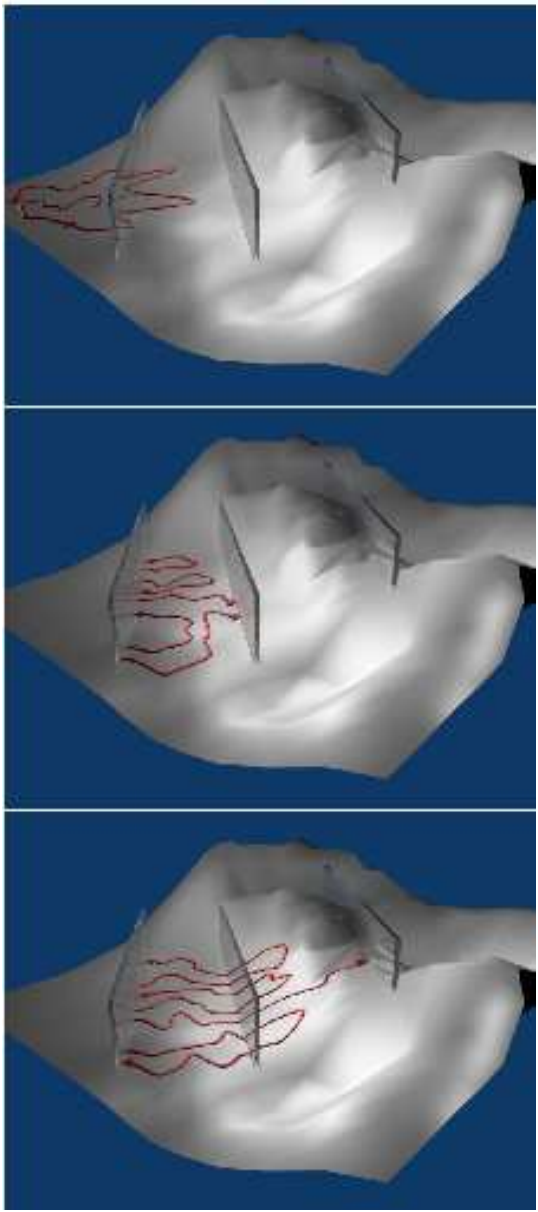}\\
   \caption{This pictures illustrates some typical trajectories on a free
energy surface for a series of TIS simulations.
The glassy plates represent the TIS interfaces. The first type of
simulation (top),
is a straightforward MD simulation
which is required to calculate the
flux $f_A$ through the first interface. The next step
is a path-sampling simulation which generates pathways that start
at $\lambda_A$ and end at either $\lambda_A$ or $\lambda_1$.
This simulation yields the result
of ${\mathcal P}_A(\lambda_1|\lambda_0)$ and is illustrated in
the middle panel.
The initial path to start this simulation can be obtained
by taking a trajectory segment from the initial MD simulation.
The bottom panel  shows the next path-sampling simulation to calculate
${\mathcal P}_A(\lambda_2|\lambda_1)$. It generates
pathways that start at $\lambda_A$ and cross $\lambda_1$ at least
once. Also here, the initial path can be obtained from the previous
simulation  by taking one of the paths that
successfully reached $\lambda_1$.
\label{figTOM}}
  \end{center}
\end{figure}

Another important point is that, in order to enter the loop,  
one needs to have a single path
that obeys the correct requirements. This can already be quite difficult 
and several techniques 
to get such a first initial path have been suggested~\cite{BolAnnu}. However, in TIS these 
initial paths are generated automatically
when the different types of simulations are consecutively performed 
(See Fig.~\ref{figTOM}).

First the MD simulation is performed to calculate $f_A$. Then, 
a series of path-sampling simulations follows to calculate
${\mathcal P}_A(\lambda_1|\lambda_0), 
{\mathcal P}_A(\lambda_2|\lambda_1),\ldots, 
{\mathcal P}_A(\lambda_n|\lambda_{n-1})$.
When these simulations are performed in this order,
each path-sampling simulation 
can obtain the
necessary initial path  
from the previous simulation (See Fig.~\ref{figTOM}). 

It is important to note that the final result, the reaction rate $k_{AB}$,
does not sensitively
depend on the positions of the outer interfaces $\lambda_A$ and $\lambda_B$
as long as they are reasonable. The number of interfaces $n$ and their
positions only influence the efficiency of the method.
It was found that the total efficiency is optimized when
for each path-simulation one out of five trajectories reaches the 
next interface~\cite{ErpBol2004,TISeff}. Hence, using some initial
trial simulations, one can adjust the number of interfaces and their position 
to satisfy this condition. The easiest way to achieve this is to use 
a slight variation of
the algorithm that is shown in Fig.~\ref{shoot}.  
Instead of stopping the integration when the trajectory
crosses $\lambda_{i+1}$ as in panel e) and k), one can continue
the trajectory until it reaches $\lambda_A$ or $\lambda_B$. 
This algorithm only requires knowing the position of $\lambda_A$, $\lambda_B$,
and $\lambda_i$. By examining the progress of the paths  along
the RC beyond $\lambda_i$, one can define the next 
interface $\lambda_{i+1}$ exactly at the point where 80 \% of the paths have
returned to $\lambda_A$.

\subsection{analysis of the reaction mechanism}
\begin{figure}[ht!]
  \begin{center}
  \includegraphics[width=8cm]{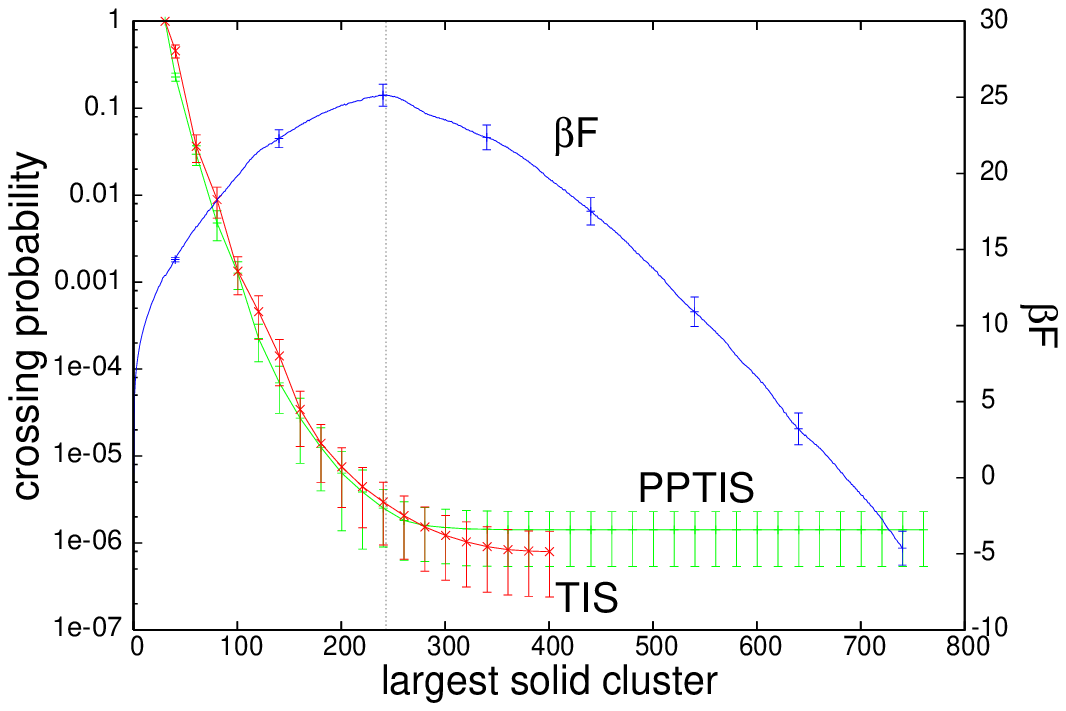}
   \caption{The crossing probability function obtained by TIS and
partial path TIS (PPTIS)~\cite{MoBolErp2004}
for nucleation process of LJ particles. The simulation data are
obtained from~[\onlinecite{MoBolPRL}].
The free energy profile,
that was calculated simultaneously using
the technique of~[\onlinecite{MoErpBol}], is also shown.
The simulated system contained 10648 particles in total.
The RC was defined as the largest solid cluster in the system.
More details can be found in ~[\onlinecite{MoBolPRL}].
The final crossing probability was $P_A(\lambda_B|\lambda_A)=(7.8 \pm 5.5)
\cdot 10^{-7}$ (TIS) and $(14.1 \pm 8.7) \cdot 10^{-7}$ (PPTIS).
The crossing probability
shows a plateau at 410.5 indicating that the reaction
barrier ridge is crossed.
The free energy barrier
has a maximum at 243. After this point
about 50~\% (PPTIS) till 75~\% (TIS) of the paths still fail
to reach the reactant state. Although, the PPTIS and TIS results are
within each others error-bars, it is likely that PPTIS overestimates
the crossing probability due to the imprecise choice of RC~\cite{TISeff}.
\label{PcrossLJ}}
  \end{center}
\end{figure}

When the complete series of simulations is finished, the reaction rate 
follows simply from Eqs.~(\ref{kTIS},\ref{kTIS2}). In addition to this,
the ensemble of pathways can be analyzed which can yield valuable information 
about the reaction mechanism. In this respect, the TIS path-ensembles 
might actually prove to be more useful than the ones
obtained by the original TPS method. As each ensemble contains 
the correct ratio of paths progressing
upto a certain level, but then either return or make a little step further, 
one can try to understand the characteristic 
differences between the 'successful' and unsuccessful' pathways. 
In contrast, the TPS method aims to generate 
successful trajectories only. 
One of the properties that can improve understanding of mechanisms
is the overall crossing probability function. This represents a sort of 
path survival probability along the RC.
This function equals 1 at $\lambda_0$ and $P_A(\lambda_B|\lambda_A)$ 
at $\lambda_B$. In transition,
this function is monotonically decreasing and terminates in a horizontal 
plateau when the barrier ridge is crossed completely.
This function could be considered as a dynamical equivalent of 
the free energy profile along the RC.
In Fig.~\ref{PcrossLJ} the overall crossing probability function 
is depicted together with the free energy profile 
obtained from a nucleation process of Lennard-Jones particles.

The results were obtained from from Ref.~[\onlinecite{MoBolPRL}]. 
Interestingly, after the 
maximum of free energy barrier is crossed (cluster size 243), 
the majority of the trajectories ($\sim 75 \%$) 
still fail to reach the reactant state.
This effect might be partly due to diffusive motion, 
but is most likely an effect of an improperly chosen RC.
This shows that the projection on a single 
RC using
static free energy calculations 
can be misleading. 
Neither the height of the barrier nor the position of the TS dividing surface 
have to reflect the actual height and position of the reaction barrier.
Indeed, Moroni \emph{et al.} found that a 'good RC' should at least incorporate 
one more important quantity which is the crystallinity of the cluster. 
Small clusters $<243$ with a high crystallinity were found to 
grow further easily, while large clusters with less structure were unstable
and broke-up into smaller pieces. 

\subsection{recent and future developments and applications}
Besides an efficient algorithm for the calculation of reaction rates,
the TIS method has provided a new mathematical framework 
to describe rare events. Recent new simulation techniques have exploited
this TIS theory. For example, for diffusive barrier crossings, where transition
paths become very long, the partial path TIS (PPTIS) method was 
devised~\cite{MoBolErp2004}. Here, using the assumption of memory loss,
much shorter paths are generated after which the overall crossing probability
can be reconstructed by a recursive formulation. The forward flux sampling
technique~\cite{FFS,FFS2} is basically the same as TIS, but the way to generate pathways is different. In this approach, the endpoints of all the pathways 
successfully reaching 
the next interface are stored and starting from each point a set of new pathways
is generated in the next simulation. 
The main advantage is that the FFS scheme supplies 
a route to handle non-equilibrium systems. A disadvantage
is that this only works
for stochastic dynamics and will always yield much stronger correlations 
between the generated pathways and the different path ensembles
even for the pure Brownian dynamics case.
Another drawback of PPTIS and FFS methods is that they 
do not possess the same 
RC insensitivity as TIS~\cite{TISeff}.
If necessary,  even a fully RC free approach is possible as 
was suggested in~[\onlinecite{ErpBol2004}]
using the TIS pathlength as a transition parameter. A nice feature of this
approach is that it
does not require to specify a specific product state.
Combinations with Configurational Bias MC~\cite{ErpBol2004,FFS2} 
and path swapping techniques~\cite{ErpBol2004,ErpInPr} may also yield
promising advances for the computational efficiency.

The TIS and its variations have been applied to various systems ranging from
simple test-systems~\cite{ErpMoBol2003,MoBolErp2004}, 
nucleation~\cite{MoBolPRL,Valer,Sear}, 
protein folding~\cite{bolhuisPNAS,pfold2},
biochemical networks~\cite{FFS}, 
driven polymer translocation through pores~\cite{FFS2}, micelle 
formation~\cite{thesisRene}, 
\emph{ab initio} simulation of chemical reactions~\cite{titusthesis}
and DNA denaturation~\cite{ErpInPr}.

The TIS technique can open many possible avenues in the field of zeolite
formation simulations at several stages of the process. 
For instance, the first elementary step to Si polymerization is
the condensation reaction 2Si(OH)$_4 \rightarrow$ Si$_2$O(OH)$_6$+H$_2$O.
This has been studied by \emph{ab initio} static analysis including
implicit solvent~\cite{Pereira3}. 
This study has revealed two possible reaction mechanisms.
Such a system is small enough to
be treated by \emph{ab initio} MD~\cite{marx,cp} including explicit solvent 
molecules. The TIS method could hence give valuable insight which 
reaction mechanism dominates when
dynamics and explicit solvent is taken into account.
Classical reactive forcefields and rare event methods,
such
as TIS and PPTIS, should 
make it possible
to simulate the dynamics of Si polymerization at much lower temperatures 
than hitherto was possible~\cite{RaoGelb}. 
This allows to study this process at conditions that are much closer to the 
experimental situation.
Moreover, by a right construction of the interfaces, the TIS method allows
to focus on reaction mechanisms and rates of some very specific
polymerization reactions, for instance the formation of the 
Si-30/33 precursor~\cite{nanoslab2}.

With the development of lattice
and KMC models, the study of the next stages of nucleation and zeolite 
growth also come into reach. The combination of KMC and path sampling
is a promising yet unexplored territory. Despite the enormous
long simulation periods that can be achieved by KMC, the expectation time to 
form a critical nucleus starting from a disorder solution is generally still
out of reach. Therefore, most KMC studies have concentrated 
on growth rather than nucleation. Hence, the study of zeolite nucleation might 
benefit significantly using combined KMC and path sampling techniques.

\section{Summary}
We have reviewed the TIS method, its variations and its possible 
applications for the theoretical study of zeolite synthesis.
The TIS method is a an elegant approach circumventing the timescale problem
not by speeding up the dynamics of the system itself, but by concentrating
on the short time trajectories which are of interest without using 
any approximation. 
TIS allows to overcome reaction barriers by a sequence of simulation series.
It is important to realize that the barrier crossing event is not 
enhanced due to some artifical force but only due to the MC acception/rejection
steps that include the interface crossing condition.
Hence, each trajectory in the
TIS path ensembles  satisfy the correct dynamics 
on the true potential energy surface. 
This makes the method fundamentally different from, for instance, 
the metadynamics~\cite{LP02} approach.
The TIS method makes use of the fact that 
the time needed to actually cross the barrier, the transition time, is much
shorter than relaxation time $k_{AB}^{-1}$, which is the time wherein one 
can expect 
a reactive event from an arbitrary point within the reactant well.

TIS can be combined with any type of dynamics such as \emph{ab initio} MD,
Langevin, pure Brownian motion, classical MD and KMC.
A requirement for application of this method is that the simulation of
short trajectories can occur sufficiently fast.
This limits the size of the systems which can be studied,
ranging from several molecules for \emph{ab initio} dynamics to several 
thousand molecules for MD, and even larger assemblies for KMC simulations. 

Still, substantial work has to be done
before fully realistic modeling of zeolite synthesis is our reach. 
An important requirement is the development of more accurate reactive force 
fields that can describe chemical events within the environment
of solvent and template molecules. 
Recently, a more systematic approach for this development was 
suggested~\cite{Toon}.
Even though, the lattice and KMC models are making substantial steps forward, 
inclusion of solvent effects in a lattice-type models has proven to be
a difficult problem that has not yet been solved. 
To conclude, the simulation methods have made prodigious advancements
in recent years 
and might ultimately give answers to important questions regarding
zeolite synthesis, that can not be unambiguously accessed 
by experimental techniques.
The TIS methods can help in obtaining dynamical 
information for the crucial but rare reaction steps in the zeolite process.
In the near future, we are going to explore 
the application of TIS for the study of zeolite genesis.

\begin{acknowledgements}
This work was sponsored by the Flemish Government
via a concerted research action (GOA), and the Belgian Government through
the IAP-PAI network. 
T.C. acknowledges the Institute for the Promotion of Innovation through Science
and Technology in Flanders (IWT-Vlaanderen) for a Ph.D. scholarship. 
\end{acknowledgements}

\bibliographystyle{pccp-bibtex}

\end{document}